\documentclass[preprint,review,10pt]{elsarticle}  

\usepackage{amssymb}
\usepackage{bm} 
\usepackage{amsbsy} 
\usepackage{amsmath} 
\usepackage{amsmath, amsthm}
\usepackage[table]{xcolor}
\usepackage{booktabs}
\usepackage{amssymb}
\usepackage{bm} 
\usepackage{dsfont} 
\usepackage{amsbsy} 
\usepackage{amsmath} 
\usepackage{amsmath, amsthm}
\usepackage[table]{xcolor}
\usepackage{booktabs}
\usepackage{graphicx}
\newtheorem{thm}{Theorem}

\newdefinition{rmk}{Remark}
\newdefinition{asp}{Assumption}
\newproof{pf}{Proof}
\newproof{pot}{Proof of Theorem \ref{thm2}}
\newdefinition{dfn}{Definition}
\newdefinition{cor}{Corollary}

\newcommand{\E}[1]{\mathbb{E}\left[ #1 \right]}

\def\bs{{\mathbf{s}}}
\def\bf{{\mathbf{f}}}
\def\bg{{\mathbf{g}}}
\def\bh{{\mathbf{h}}}
\def\bx{{\mathbf{x}}}
\def\bd{{\mathbf{d}}}
\def\bX{{\mathbf{X}}}
\def\be{{\mathbf{e}}}
\def\bu{{\mathbf{u}}}

\def\bv{{\mathbf{v}}}
\def\bw{{\mathbf{w}}}
\def\bA{{\mathbf{A}}}
\def\bB{{\mathbf{B}}}
\def\bC{{\mathbf{C}}}
\def\bF{{\mathbf{F}}}

\def\bP{{\mathbf{P}}}
\def\wbh{{\mathbf{h}}}
\def\wbg{{\mathbf{g}}}
\def\bC{{\mathbf{C}}}
\def\bU{{\mathbf{U}}}
\def\bD{{\mathbf{D}}}
\def\bZ{{\mathbf{Z}}}

\def\BF{\bm{\mathcal{F}}}
\def\BP{\bm{\mathcal{P}}}
\def\BM{\bm{\mathcal{M}}}
\def\BN{\bm{\mathcal{N}}}
\def\BH{\bm{\mathcal{H}}}

\def\vec{{\mathsf{vec}}}
\def\E{{\mathbb{E}}}
\def\bI{{\mathbf{I}}}

\def\T{{\mathsf{T}}}
\def\H{{\mathsf{H}}}
\def\E{{\mathsf{E}}}

\def\bSigma{{\bm{\Sigma}}}
\def\bsigma{{\bm{\sigma}}}
\def\bgamma{{\bm{\gamma}}}

\def\Tr{{\mathsf{Tr}}}

 \biboptions{numbers,square,comma}

\begin{document}

\begin{frontmatter}



\title{Analysis of incremental augmented affine projection algorithm for distributed estimation of complex signals}

 \author[label1]{Azam Khalili}
 \address[label1]{Department of Electrical Engineering, Malayer University, Malayer, 65719-95863, Iran}

\author[label2]{Wael M. Bazzi}
 \address[label2]{Electrical and Computer Engineering Department, American University in Dubai, Dubai, United Arab  Emirates}
 
\author[label1]{Amir~Rastegarnia\corref{cor1}}
\cortext[cor1]{corresponding author, Email: a\_rastegar@ieee.org}

\begin{abstract}
This paper considers the problem of distributed estimation in an incremental network when the measurements taken by the node follow a widely linear model. The proposed algorithm which we refer to it as incremental augmented affine projection algorithm (incAAPA) utilizes the full
second order statistical information in the complex domain. Moreover, it exploits spatio-temporal diversity to improve the estimation performance.
We derive steady-state performance metric of the incAAPA in terms of the mean-square deviation (MSD). We further derive sufficient
conditions to ensure mean-square convergence. Our analysis illustrate that the proposed algorithm is able to process both second order circular (proper) and noncircular (improper) signals. The validity of the theoretical results and the good performance of the proposed algorithm are demonstrated by several computer simulations. 
\end{abstract}


\end{frontmatter}


\section{Introduction}
\label{sec:1}
In many practical applications the ultimate goal is to estimate an unknown parameter of interest by using observations acquired by spatially distributed nodes \cite{barb13,gia06}. Different solutions (algorithms) have been introduced in the literature to solve the decentralized estimation problem. In some algorithms (known as consensus methods) such as \cite{sch08a, sch08b} the nodes collect all the data first, perform local estimation and then interact iteratively with their neighbors. Networks that rely on in-network processing at each node while allowing the node to learn with new observations are known as adaptive networks \cite{sayed13b, sayed14}. An adaptive network consists of a collection of spatially distributed nodes that are able to communicate with each other through a topology. Two major class of adaptive networks, based on the network topology are incremental networks
\cite{sayed06,lopes07,tak08,Li08,li10,lopes10,ras10,catt11b,Saeed12,shams13,Berb13,Li14,bog13} or diffusion algorithms \cite{lopes08,catt08,catt09,tak09,catt10a,Ghar13,Di14,chen12,pen13,arab14,wen14}. In the incremental algorithms, a cyclic path through the network is established, and nodes communicate with neighbors within this path. In adaptive diffusion implementations, information is processed locally at the nodes and then diffused in real-time across the network. 

\subsection{Motivation for current work}
An important factor in learning abilities of adaptive networks is the adaptive filter that is embedded at the nodes. The mentioned adaptive networks use different types of adaptive filters such as leas meant-square (LMS) \cite{lopes07,catt11b}, recursive least-squares (RLS) \cite{sayed06,catt08}, affine projection algorithm (APA) \cite{Li08,li10} and normalized least mean squares (NLMS) \cite{baq13}. 
The LMS algorithm is a popular choice due to its stability and  low complexity. On the other hand, the algorithms based on the least squares criterion  such as the RLS algorithm, converge faster. However, they suffer from the high computational complexity issue \cite{sayed08}. The affine projection algorithm has the   benefits of both approaches that is stability, low computational complexity and fast convergence \cite{sayed08, xia10}. 

The APA was originally introduced for real-value signals \cite{Ben96,Arab12}. The complex domain provides a natural processing framework for 
signals with intensity and direction components \cite{man09,adali11}. Statistical signal processing in $\mathbb{C}$ has traditionally been viewed as a straightforward extension of the corresponding algorithms in the real domain $\mathbb{R}$. However, recent developments in augmented complex statistics show that they do not make full use of the algebraic structure of the complex domain \cite{man09}. For example,
it was shown that the covariance matrix is not sufficient to model the statistics of noncircular signals and it is necessary to introduce the pseudocovariance
matrix to fully capture the relation between the real and imaginary components
of random vectors. It is also shown that the standard linear model is only sufficient for
modelling 'proper' signals, whereas an optimal model for 'improper' signals is provided
by a widely linear model \cite{javid07,adali12}.

\subsection{Contributions}
The aforementioned adaptive networks rely on linear model and signal processing in $\mathbb{R}$ domain. 
In this paper, we develop and analyze an incremental adaptive network based on the widely linear model which employs the affine projection algorithm as learning entity to process the complex-valued signals. The proposed algorithm, which we refer to as incremental augmented affine projection algorithm (incAAPA), exploits spatio-temporal diversity to improve the estimation performance. Moreover, at the same time, it employs augmented complex statistics  which it turn enables it to process both types of proper and improper signals. To analyze the steady-state performance of the IncAAPA  by resorting to the weighted energy conservation approach and derive a closed-form expression, in terms of the mean-square deviation (MSD) that explains how the algorithm performs in the steady-state. 
We further derive the stability bound of the proposed algorithm to ensure mean-square convergence.
We use  different synthetic benchmark signals including circular complex-valued signal, the noncircular complex-valued signal and noncircular chaotic Ikeda map signal in our simulations to evaluate the performance of the proposed algorithm. Simulation results validate the theoretical results and reveal the advantage of using the proposed algorithm for processing of both second order circular (proper) and noncircular (improper) signals.

\subsection{Paper organization and notation}
The remainder of this paper is organized as follows. In Section \ref{sec:2}, we formulate the estimation problem and derive the proposed algorithm.
In Section \ref{sec:3}, performance analysis of the incAAPA algorithm is provided. In Section \ref{sec:4}, we present simulation results to verify our theoretical analysis, and we conclude in Section \ref{sec:5}.

We adopt small boldface letters for vectors and bold capital letters for matrices. The other symbols used in this paper are listed in
Table. \ref{tab:1}.

\begin{table*}
	\caption{Symbols and their descriptions.}
	\scalebox{0.8}{
\begin{tabular}{l l }
\toprule
$x$, $\bx$, $\bX$  & Scalar, column vector, matrix; \\
$\Tr[\bA]$          &  Trace of matrix $\bA$,  \\
$(\cdot)^{\T}$,  $(\cdot)^{\H}$, $(\cdot)^{*}$                    & Transpose, Hermitian transpose, Complex conjugation   \\
$\E[\cdot]$			&	Statistical expectation operator;  \\
$\otimes $								& Kronecker product   \\
$\Re\{\}$,  $\Im\{\}$  &   Real, Imaginary parts of a complex number \\
$\vec\{\cdot \}$ &  This notation will be used in two ways: $\bx  = {\vec}\{ \bSigma \} $ is a $M^2  \times 1$ column vector\\ 
                       & whose entries are formed by stacking the successive columns of an $M \times M$ matrix \\
                       & on top of each other, and $\bX  = {\vec}\{ \bx \}$ is a matrix whose entries are recovered from $\bx$. \\
 $\| \bu \|_{\bSigma}^2=\bu^H \bSigma \bu$  &  the weighted square norm of $\bu$. \\
 $\lambda_{\max}(\bA)$ & The largest eigenvalue of matrix $\bA$.\\ 
\bottomrule
\end{tabular}}
\label{tab:1}
\end{table*}

\section{Proposed Algorithm}\label{sec:2}
\subsection{The estimation problem}
\begin{figure}[t]
\centering 
\includegraphics [width=8cm] {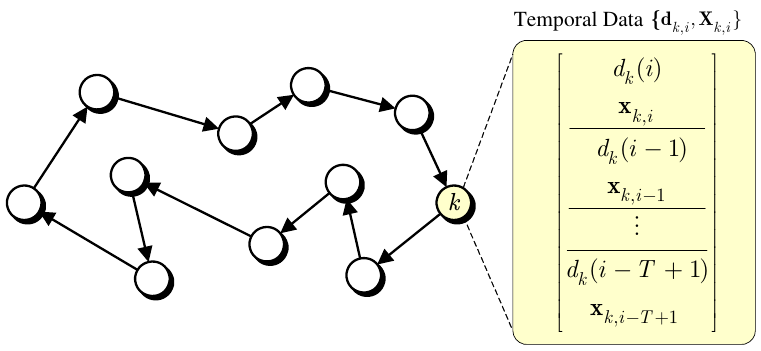}  
\centering\caption{An incremental network with $N$ nodes.}
\label{fig:1}
\end{figure}

Let denote by set $\mathcal{K}=\{1,2,\cdots,N\}$ a network with $N$ nodes which communicate according to the incremental
protocol. This protocol require a Hamiltonian cycle through which local information are sequentially circulated from node to node (See Fig. \ref{fig:1}).
At any time $i$, node $k$ measures data $\{d_k(i), \bx_{k,i}\}$ where $d_k(i)\in \mathbb{C}$  is the desired signal and 
$\bx_{k,i}=[x_{k,i}(1), x_{k,i}(2), \cdots, x_{k,i}(L)]^{\T} \in \mathbb{C}^{L \times 1}$ denotes the input vector at node $k$. we consider the following assumptions on our data model.
\begin{asp} \label{ass:1}
The desired signal $d_k(i)$ and the input vector $\bx_{k,i}$ are related through a widely linear model as \cite{man09}
\begin{equation} \label{model}
{d_k}(i) = \bx_{k,i}^{\T}{\bh^o} + \bx_{k,i}^{\H}{\bg^o} + {v_k}(i),
\end{equation}
where $\{\bh^o, \bg^o\} \in \mathbb{R}^{L \times 1}$ are the unknown parameters, and $v_k(i)\in \mathbb{C}$ denotes samples of doubly white noise with variance $\sigma_{v,k}^2$.  The measurement noises $\{v_k(i)\}$ for all nodes $k=1,\ldots,N$, and all observation times $i \geq 1$, are independent of each other and the regression vectors $\bx_{k,i}$.
\end{asp} 
\begin{asp} \label{ass:2}
For $k=1,\ldots,N$ and $i \geq 1$, the regression vectors $\bx_{k,i}$ are independent over node indices $k$ and observation times $i$. 
\end{asp}
The measurements model \eqref{model} appears in many practical applications, such as the frequency estimation problem in three-phase power systems \cite{kanna14}, and in  simultaneous modelling and forecasting of wind signals \cite{yili11,Man09b}. 
\begin{dfn} \emph{(Noncircular signals)}
To define the noncircular (or improper) signals, let define the augmented vector $\bx_{k,i}^{a}$ as
\[
\bx_{k,i}^{a} \triangleq  
\left[ {\begin{array}{*{20}{c}}
   \bx_{k,i} \\
   \bx_{k,i}^*  \\
\end{array}} \right],\ \ (2L \times 1)
\]
The covariance matrix $\bC_{\bx \bx,k}^{a}$ for the augmented vector $\bx_{k,i}^{a}$ is given by
\begin{equation}
\bC_{\bx \bx,k}^a = \E[\bx_{k,i}^{a} \bx_{k,i}^{a \H}]
=
\left[ {\begin{array}{*{20}{c}}
   \bC_{\bx \bx,k} &  \bP_{\bx \bx,k}  \\
   \bP_{\bx \bx,k}^ *  & \bC_{\bx \bx,k}^ *   \\
\end{array}} \right]
\end{equation}
when the pseudo-covariance matrix $\bP_{\bx \bx,k}=\mathbf{0}$, the complex random vector is called circular (proper). 
\end{dfn}

\subsection{Algorithm derivation}
To develop the proposed algorithm we denote by $\{\bd_{k,i}, \bX_{k,i}\}$ the $T$ most recent observations at node $k$ as
\begin{align} \label{mrdata}
 \bX_{k,i} &= [\bx_{k,i },...,\bx_{k,i- T + 1}],\ \  (L\times T)   \\
 \bd_{k,i} &= [{d_k}(i ),...,{d_k}(i- T + 1)]^{\T},\ \  (T\times 1) 
\end{align} 
Our objective is to develop a distributed and adaptive algorithm that is able to estimate the unknown parameters $\{\bh^o,\bg^o\}$ at every node $k\in \mathcal{K}$. To achieve this objective, we formulate the desired estimation problem as the following constrained optimization problem which is based on the minimum disturbance \cite{cast13} 
\begin{align} \label{consf}
\mathrm{minimize} & \hspace{0.2cm}
f(\wbh_i,\wbg_i) \triangleq \Big({\| \wbh_{i} - {\wbh_{i-1}} \|^2} + {\| {\wbg_{i}} - {\wbg_{i-1 }} \|^2}\Big) \\
\mathrm{subject\ to}  & \hspace{0.2cm} \sum_{k=1}^{K}{\mathop{\Re}\nolimits} \left\{ {{({\bd_{k,i}} - \bX_{k,i}^{\T} {\wbh_{i}} - \bX_{k,i}^{\H}{\wbg_{i}})}^{\H}}\right\}=0  
\nonumber
\end{align}
In the following theorem we discuss the solution of the problem \eqref{consf}.
\begin{thm}
Let $\bh_{k,i}$ and $\bg_{k,i}$ denote the local estimates of $\bh^o$ and $\bg^o$ at node $k$ at time $i$ respectively. The optimal wights $\{\bh^o,\bg^o\}$ can be recursively estimated at every node $k$ via the following update equations 
\begin{align} 
 {\bh_{k,i}} &= {\bh_{k - 1,i}} + {\mu _k}\bX_{k,i}^ *{\bB_{k,i}} {\be_{k,i}}    \label{hupdate} \\
 {\bg_{k,i}} &= {\bg_{k - 1,i}} + {\mu _k}{\bX_{k,i}} {\bB_{k,i}} {\be_{k,i}}    \label{gupdate}
\end{align}
where 
\begin{align} 
{\bA_{k,i}} & \triangleq \bX_{k,i}^{\H}{\bX_{k,i}} + \bX_{k,i}^{\T} \bX_{k,i}^ *   \\
{\bB_{k,i}} & \triangleq (\bA_{k,i}+ \delta \bI)^{-1} 
\end{align}
Moreover, $\delta >0$ is the regularisation parameter, $\mu>0$ is the step-size parameter and $\be_{k,i}$ is the local error signal which is defined as
\begin{equation} \label{defe}
\be_{k,i}\triangleq {\bd_{k,i}}-\bX_{k,i}^{\T} \bh_{k-1,i} -\bX_{k,i}^{\H}{\bg_{k-1,i}} 
\end{equation}
\end{thm}
\begin{pf}
See \cite{amir2014}.
\end{pf}
\begin{rmk}
It must be noted that a non-cooperative solution for the problem \eqref{consf} based on the affine projection algorithm is given by
\begin{align}  \label{nocoop}
 {\bh_{k,i}} &= {\bh_{k,i - 1}} + {\mu _k}\bX_{k,i}^ *{\bB_{k,i}} {\be_{k,i}}     \\
 {\bg_{k,i}} &= {\bg_{k,i - 1}} + {\mu _k}{\bX_{k,i}} {\bB_{k,i}} {\be_{k,i}}    
\end{align}
where reveals that in this solution every node uses solely its local information.
\end{rmk}
An important question now is, how well does the adaptive incremental solution \eqref{hupdate} and \eqref{gupdate} perform? That is, how close does each 
$\{\bh_{k,i},\bg_{k,i}\}$ get to the desired solutions $\{\bh^o,\bg^o\}$ as time evolves? In the next section, we provide a framework for studying the performance of such network by examining the flow of energy through the network both in time and space.

\section{Performance Analysis} \label{sec:3}
In order to study the performance of the IncAAPA algorithm, we extend the weighted energy conservation argument  for the stand alone adaptive filters \cite{sayed08} to the case of incremental networks. we evaluate the  steady-state performances at each individual node in terms of mean-square deviation (MSD). 
To begin the analysis we define the following weight error vectors as
\begin{equation} \label{wevec}
{{\tilde \bg}_{k,i}} = {\bg_0} - {\bg_{k,i}},\ \ {{\tilde \bh}_{k,i}} = {\bh_0} - {\bh_{k,i}}
\end{equation}
using \eqref{wevec}, the update equations in \eqref{hupdate} and \eqref{gupdate} can be rewritten in terms of the weight error vectors as
\begin{align} 
{{\tilde \bh}_{k,i}} &= {{\tilde \bh}_{k - 1,i}} - {\mu _k} \bX_{k,i}^ * \bB_{k,i}{\be_{k,i}}  \label{tildh}   \\
{{\tilde \bg}_{k,i}} &= {{\tilde \bg}_{k - 1,i}} - {\mu _k}{\bX_{k,i}}\bB_{k,i} {\be_{k,i}}    \label{tildg} 
\end{align}
Multiplying both sides of \eqref{tildh} and \eqref{tildg} by $\bX_{k,i}^{\T}\bSigma_{k}$  and $\bX_{k,i}^{\H} \bSigma_{k}$ respectively gives
\begin{align} 
\bX_{k,i}^{\T}\bSigma_{k}{{\tilde \bh}_{k,i}} &= \bX_{k,i}^{\T}\bSigma_{k}{{\tilde \bh}_{k - 1,i}} - {\mu _k} \bX_{k,i}^{\T} \bSigma_{k} \bX_{k,i}^ * \bB_{k,i}{\be_{k,i}}  \label{addte1}  \\
\bX_{k,i}^{\H}\bSigma_{k}{{\tilde \bg}_{k,i}} &= \bX_{k,i}^{\H}\bSigma_{k}{{\tilde \bg}_{k - 1,i}}- {\mu _k} \bX_{k,i}^{\H}\bSigma_{k}{\bX_{k,i}}\bB_{k,i} {\be_{k,i}}   \label{addte2}
\end{align}
The weighted a-posteriori and weighted a-priori error vectors $\{\be^{\bSigma_{k}}_{p,k},\be_{a,k}^{\bSigma_{k}} \}$ are introduced as
\begin{align} 
\be^{\bSigma_{k}}_{p,k} &= \bX_{k,i}^T \bSigma_{k} {{\tilde \bh}_{k,i}} + \bX_{k,i}^H \bSigma_{k}{{\tilde \bg}_{k,i}}   					\label{post}  \\
\be^{\bSigma_{k}}_{a,k} &= \bX_{k,i}^T \bSigma_{k} {{\tilde \bh}_{k - 1,i}} + \bX_{k,i}^H \bSigma_{k} {{\tilde \bg}_{k - 1,i}}    \label{prio}
\end{align}
If we add \eqref{addte1} and \eqref{addte2} and use the definitions of $\be^{\bSigma_{k}}_{p,k}$ and $\be^{\bSigma_{k}}_{a,k}$ we get
\begin{align} \label{pvera}
\be^{\bSigma_{k}}_{p,k} &= \be^{\bSigma_{k}}_{a,k} - {\mu _k} \bF_{k,i}\bB_{k,i}{\be_{k,i}}  
\end{align}
where
\begin{align} 
{\bF_{k,i}} & \triangleq \bX_{k,i}^{\H}\bSigma_{k}{\bX_{k,i}} + \bX_{k,i}^{\T}\bSigma_{k} \bX_{k,i}^ * 
\end{align}
using \eqref{pvera}, we can be expressed ${\be_{k,i}}$ as 
\begin{equation}  \label{edef}
{\be_{k,i}} = \frac{1}{\mu_k}\bB_{k,1}^{-1} \bF_{k,i}^{-1}\big(\be^{\bSigma_{k}}_{a,k} - \be^{\bSigma_{k}}_{p,k}\big)
\end{equation}
Now we define the augmented quantities $\tilde{\bw}_{k,i}$ and $\bU_{k,i}$ as
\begin{equation}  \label{wudef}
\tilde{\bw}_{k,i} \triangleq  
\left[ {\begin{array}{*{20}{c}}
   \tilde{\bh}_{k,i} \\
   \tilde{\bg}_{k,i}  \\
\end{array}} \right]_{2L \times 1},\ \ 
\bU_{k,i} \triangleq  
\left[ {\begin{array}{*{20}{c}}
   {\bX}^*_{k,i} \\
   {\bX}_{k,i}  \\
\end{array}} \right]_{2L \times T},
\end{equation}
\begin{cor}
Using the definition in \eqref{wudef} we have 
\begin{subequations}
\begin{align}
\be^{\bSigma_{k}}_{p,k}&=\bU_{k,i}^{\H} \bSigma_{k}\tilde{\bw}_{k,i},    \\
 \be^{\bSigma_{k}}_{a,k}&=\bU_{k,i}^{\H} \bSigma_{k}\tilde{\bw}_{k-1,i}   \\
\bF_{k,i}&=\bU_{k,i}^{\H} \bSigma_{k} \bU_{k,i}   \label{fnewd}
\end{align}
\end{subequations}
\end{cor}
This definitions in \eqref{wudef} allows us to rewrite equations \eqref{tildh} and\eqref{tildg} as the following equivalent form 
\begin{align} \label{wtildup}
\tilde{\bw}_{k,i}&=\tilde{\bw}_{k-1,i}-\bU_{k,i}\bF_{k,i}^{-1}\big(\be^{\bSigma_{k}}_{a,k} - \be^{\bSigma_{k}}_{p,k}\big)
\end{align}
To derive the desired steady-state performance metric, we need to derive the mean-square variance relation that explains how $\{\bh_{k,i},\bg_{k,i}\}$ evolves in time. Thus, we assert the following theorem on the mean-square variance relation of the incAAPA.
\begin{thm} \label{thm:2}
The mean-square variance relation of the incAAPA is given by the following expression.
\begin{equation}       \label{36x}
\E \big[\|\tilde{\bw}_{k,i}\|_{\bSigma_{k}}^2\big] =\E \big[\|\tilde{\bw}_{k-1,i}\|_{\bSigma'_{k}}^2\big]
+\mu_k^2 \E\big[\bv_{k,i}^{\H}\bD_{k,i}\bSigma_{k}\bD_{k,i}\bv_{k,i}\big]
\end{equation}
\begin{align} \label{37x}
\bSigma'_k&=\bSigma_k-\mu_k \bSigma_k \E [\bD_{k,i}]-\mu_k \E [\bD_{k,i}] \bSigma_k+\mu_k^2 \E [\bD_{k,i} \bSigma_k \bD_{k,i}]
\end{align} 
\end{thm}
\begin{pf}
See Appendix B for details.
\end{pf}
To derive the desired metric we apply the $\vec(\cdot)$ operator on both sides of \eqref{36x} and \eqref{37x} and get\footnote{For any matrices of compatible dimensions $\{\bZ_1, \bZ_2, \bSigma_k\}$, it holds that
\begin{align} 
\vec\{\bZ_1 \bSigma_k \bZ_2\}&=(\bZ_2^{\T} \otimes \bZ_1) \vec\{\bsigma_k\}   \nonumber  
\end{align}
}
\begin{align} 
\vec\{\bSigma_k \E [\bD_{k,i}]\} &=\left(\E[\bD_{k,i}^{\T}]\otimes \bI \right)\bsigma_k    \label{39a}   \\
\vec\{\E [\bD_{k,i}]\bSigma_k \} &= \left(\bI \otimes \E[\bD_{k,i}] \right)\bsigma_k       \label{39b} \\
\vec\{\E [\bD_{k,i}\bSigma_k \bD_{k,i}] \} &= \left(\E[\bD_{k,i}^{\T}] \otimes \E[\bD_{k,i}] \right) \bsigma_k  \label{39c}
\end{align} 
Therefore, using \eqref{39a}-\eqref{39c}, we obtain a linear relation between the corresponding vectors $\{\bsigma'_k,\bsigma_k\}$, namely
\begin{align} \label{40}
\bsigma'_k &= \BF_k \bsigma_k
\end{align}
where $\BF_k$ is given by
\begin{align} \label{41}
\BF_k&= \bI -\mu_k \left(\E[\bD_{k,i}^{\T}] \otimes \bI \right)-\mu_k \left(\bI \otimes \E[\bD_{k,i}]\right)+\mu_k^2 \left(\E[\bD_{k,i}^{\T}]\otimes \E[\bD_{k,i}]\right)
\end{align}
The second term in right hand side of \eqref{36x} can be rewritten as 
\begin{align} \label{41b}
\bs_k &=\E\big[\bv_{k,i}^{\H}\bD_{k,i}\bSigma_{k}\bD_{k,i}\bv_{k,i}\big] \nonumber \\
      &=\Tr \bigg[\E\big[\bv_{k,i}^{\H}\bD_{k,i}\bSigma_{k}\bD_{k,i}\bv_{k,i}\big]\bigg]    \nonumber  \\
      &=\E \bigg[\Tr \big[\bv_{k,i}^{\H}\bD_{k,i}\bSigma_{k}\bD_{k,i}\bv_{k,i}\big]\bigg]   \nonumber  \\
      &=\E \bigg[\Tr \big[\bD_{k,i}\bv_{k,i}\bv_{k,i}^{\H}\bD_{k,i}\bSigma_{k}\big]\bigg]   \nonumber  \\
      &=\sigma_{v,k}^2 \bgamma_k^{\T} \bsigma_k
\end{align}
where $\bv_{k,i} = [{v_k}(i),...,{v_k}{(i - T + 1)}]^{\T}$, $\bD_{k,i}=\bU_{k,i}\bB_{k,i}\bU_{k,i}^{\H}$ and
\begin{align}      
\bgamma_k &= \vec\{\E [\bD_{k,i}^2]\}
\end{align} 
As a result, expression \eqref{36x} becomes
\begin{align}       \label{fin}
\E \big[\|\tilde{\bw}_{k,i}\|_{\bsigma_{k}}^2\big] &=\E \big[\|\tilde{\bw}_{k-1,i}\|_{\bsigma'_{k}}^2\big]
+\mu_k^2 \delta_{v,k}^2 \bgamma_k^{\T} \bsigma_k     
\end{align} 
 Observe, that \eqref{fin} is a coupled equation: it involves both $\E [\|\tilde{\bw}_{k,i} \|_{\sigma _{k} }^2]$ and $\E[\|\tilde{\bw}_{k - 1,i} \|_{\bsigma'_{k} }^2]$, i.e., information from two spatial locations. The ring topology together with the weighting matrices can be exploited to resolve this difficulty \cite{lopes07} where a similar equation is solved for incremental LMS algorithm. If we follow the steps given in \cite{lopes07} we can obtain the following equation for incAAPA algorithm
\begin{eqnarray} \label{wvsf}
 \E \bigg[\|\tilde{\bw}_{k - 1} \|_{\bsigma _{k - 1} }^2\bigg]  &=& \E \bigg[\|\tilde{\bw}_{k - 1} \|_{\BF_k  \ldots \BF_N \BF_1  \ldots \BF_{k - 1} \bsigma _{k - 1} }^2\bigg]   \nonumber \\ 
 &&  + \bs_k \BF_{k + 1}  \ldots \BF_N \BF_1  \ldots \BF_{k - 1} \bsigma _{k - 1}  \nonumber  \\
 &&  + \bs_{k + 1} \BF_{k + 2}  \ldots \BF_N \BF_1  \ldots \BF_{k - 1} \bsigma _{k - 1} \nonumber  \\
 & & + \bs_{k - 2} \BF_{k - 1} \bsigma _{k - 1}  + \bs_{k - 1} \bsigma _{k - 1}
\end{eqnarray}
 Let
\begin{equation}  \label{pai}
\BP_{k,l}= \BF_{k + l - 1}  \ldots \BF_N \BF_1  \ldots \BF_{k - 1} ,\quad l = 1,2, \ldots ,N 
\end{equation}
\begin{equation} \label{fi}
 \bf_k= \bs_k \BP_{k,2}  + \bs_{k + 1} \BP_{k,3}  +  \ldots  + \bs_{k - 2} \BP _{k,N}  + \bs_{k - 1} 
\end{equation}
where in \eqref{pai} the subscripts are all \emph{mod} $N$. Using \eqref{pai} and \eqref{fi}, we can represent \eqref{wvsf} in the following form 
\begin{equation} \label{eq:40}
\E \Big[\left\| \tilde{\bw}_{k - 1} \right\|_{(\bI - \BP_{k,1} )\bsigma _{k - 1} }^2\Big]  = \bf_k \bsigma _{k - 1}
\end{equation}
Expression \eqref{eq:40} can be exploited to evaluate the desired performance measures at node $k$. In fact, since we are free to select the weight vector $\sigma_{k-1}$, choosing $\bsigma_{k-1}=(\bI - \BP_{k,1} )^{ - 1}\vec\{\bI\}$ results in the expression for the steady-state MSD at node $k$ as follows
\begin{align}
\mathrm{MSD}_k  &=\bf_k (I - \BP_{k,1} )^{ - 1} \vec\{\bI\} \\
\end{align}
To derive the required mean-square stability condition we refer to \eqref{fin}. The recursion is stable if the matrix $\BF_k$ is stable. From \eqref{41}, we can rewrite $\BF_k$ as
\begin{equation}
\BF_k= \bI -\mu_k \BM_k +\mu_k^2 \BN_k
\end{equation}
where
\begin{align}
\BM_k&=\left(\E[\bD_{k,i}^{\T}] \otimes \bI \right)+\left(\bI \otimes \E[\bD_{k,i}]\right) \\
\BN_k&= \left(\E[\bD_{k,i}^{\T}]\otimes \E[\bD_{k,i}]\right)
\end{align}
Following the same approach in \cite{shin04}, the condition on step-size that guarantees the convergence of the proposed algorithm in the mean-square is given as
\begin{equation}
0<\mu_k<\min \left\{\frac{1}{\lambda_{\max}(\BM_k^{-1}\BN_k)},\frac{1}{\lambda_{\max}(\BH_k)}\right\}
\end{equation}
where
\begin{equation}
\BH_k = \begin{bmatrix}
       \frac{1}{2}\BM_k &  - \frac{1}{2}\BN_k           \\[0.3em]
       \bI &  \mathbf{0}           \\[0.3em]
       \end{bmatrix}
\end{equation}

\section{Simulation Results} \label{sec:4}
\begin{figure*}[t]
\centering 
\includegraphics [width=12cm]{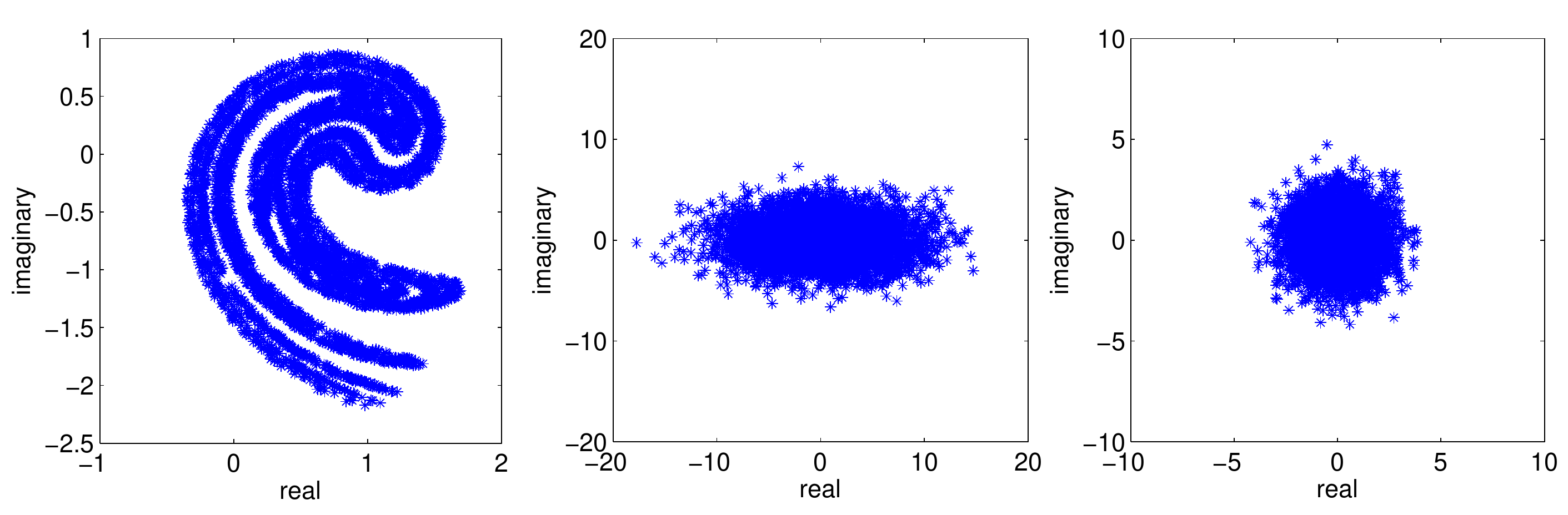} 
\centering \caption{Geometric view of the synthesized circular complex-valued  signal (left), synthesized noncircular complex-valued  signal (middle) and the nonlinear and noncircular chaotic Ikeda map signal (right).}
\label{fig:2}
\end{figure*}

\begin{figure}[t]
\centering 
\includegraphics [width=8.5cm]{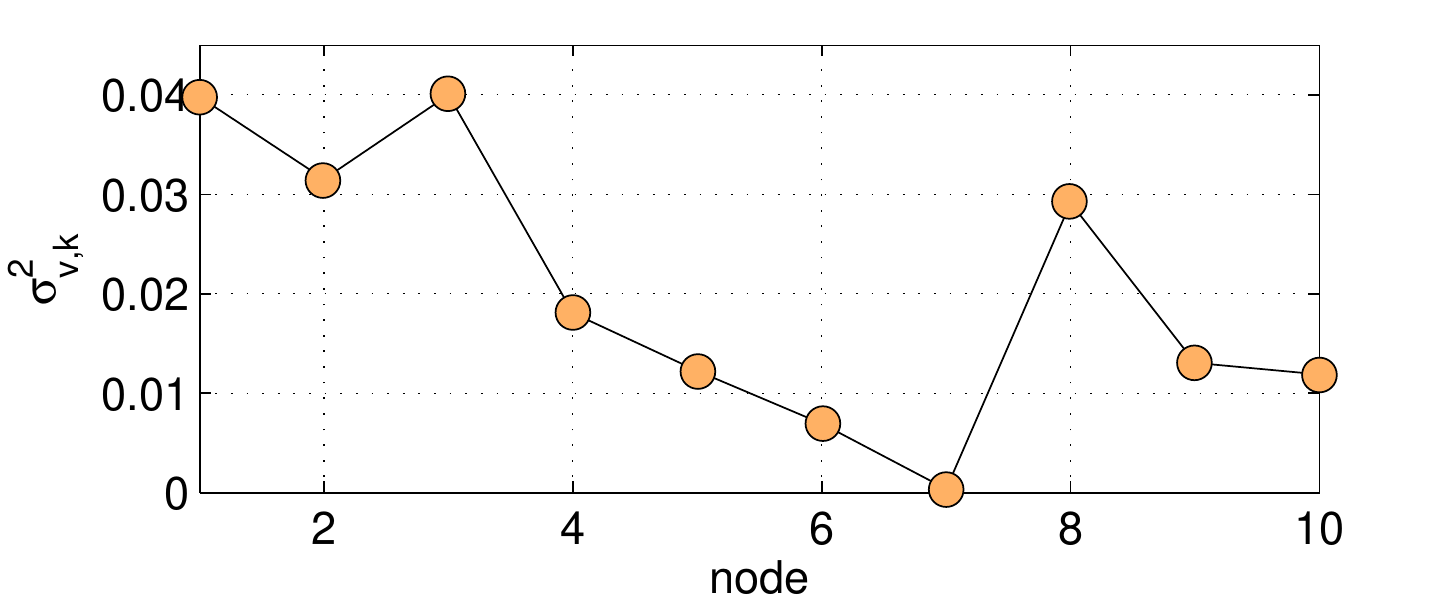} 
\centering \caption{The measurement noise profile $\sigma_{v,k}^2$.}
\label{fig:3}
\end{figure}

\begin{figure}[t]
\centering 
\includegraphics [width=8.5cm]{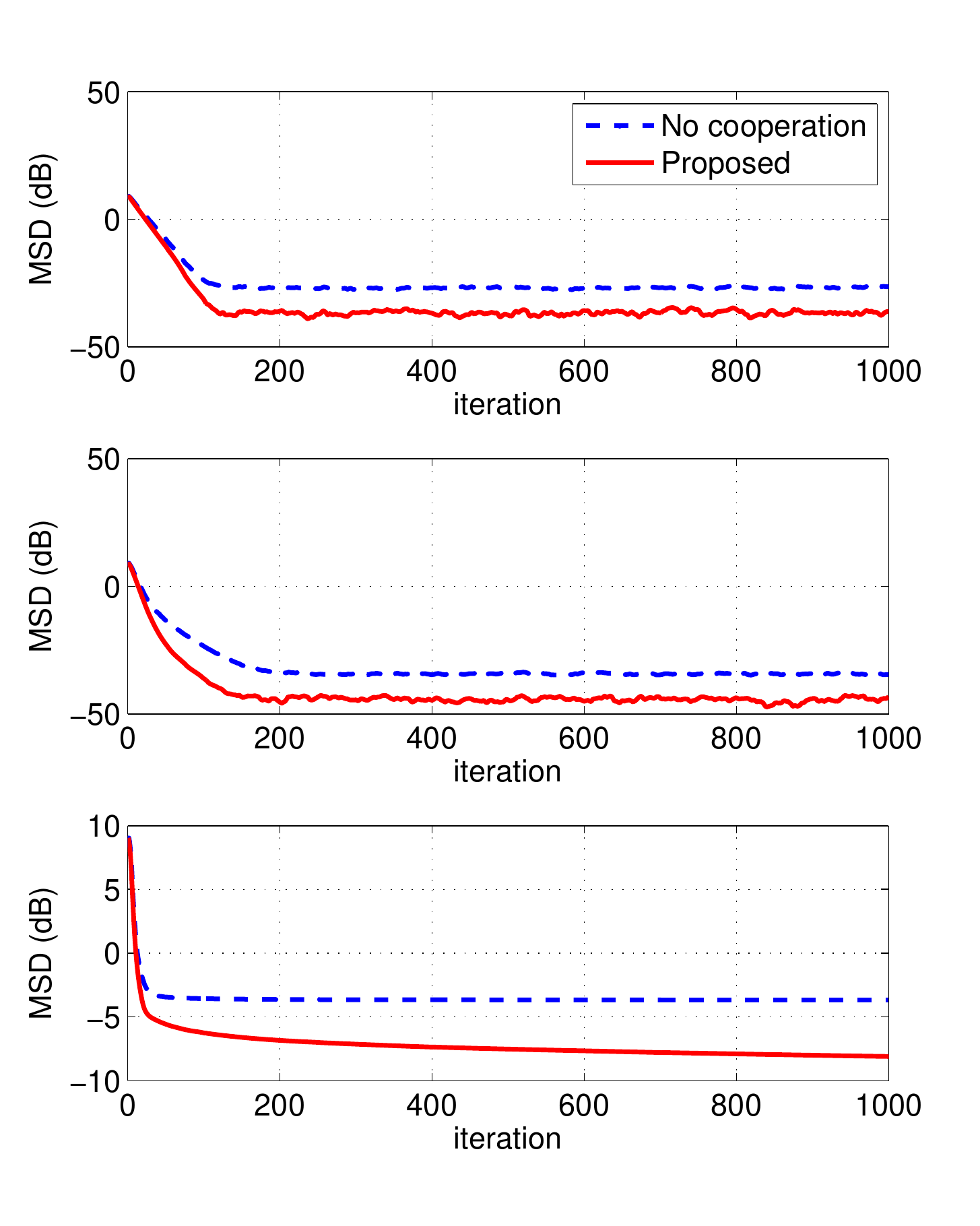} 
\centering \caption{The MSD learning curves  for non-cooperative scheme and the proposed algorithm for different signals: circular complex-valued (top), synthesized noncircular complex-valued  signal (middle) and the nonlinear and noncircular chaotic Ikeda map signal (bottom).}
\label{fig:4}
\end{figure}

In order to illustrate the performance of the proposed algorithm, in this section we present the simulation results for a network with $N=10$ nodes. We consider three different models for the input vectors  including 
\begin{itemize}
	\item The synthesized circular complex-valued  signal which is generated at every node by a AR(1) model as
	\footnote{Note that $\bx_{k,i}=[x_{k,i}(1), x_{k,i}(2), \cdots, x_{k,i}(L)]$}
	\begin{equation}
	x_{k,i}(t)=0.5x_{k,i}(t-1)+q_{k,i}
	\end{equation}
	where $q_{k,i}$ is complex valued doubly white Gaussian noise with unit variance.
	\item The synthesized noncircular complex-valued  signal which is generated at every node by a ARMA model as
	\begin{equation}
	x_{k,i}(t)=0.5x_{k,i}(t-1)+2q_{k,i}+0.5q^{*}_{k,i}+q_{k,i-1}+0.9q^{*}_{k,i-1}
	\end{equation}
	\item The noncircular chaotic Ikeda map signal which is given by
	\begin{align}
	    a(t)&=1+0.9 (a(t-1)\cos(r_k(t-1))-b(t-1) \sin(r_k(t-1))); \nonumber \\
	    b(t)&=0.9 (a(t-1)\sin(r_k(t-1))+b(t-1)\cos(r_k(t-1)));
	\end{align}
	where $x_r(t)=\Re\{x_{k,i}(t)\}$, $b(t)=\Im\{x_{k,i}(t)\}$, $x_{k,i}(t)=a(t)+j b(t)$ and 
	\[
	r_k(t)=0.4-(6/(1+a(t)^2+b(t)^2));
	\]
	\end{itemize}
Fig. \ref{fig:2} shows the scatter plots of the synthesized circular complex-valued  signal, synthesized noncircular complex-valued  signal and the nonlinear and noncircular chaotic Ikeda map signal.
We also consider the measurement model \eqref{model} with $L=4$ and $\bh^o=\bg^o=[1\ 1\ 1\ 1]^{\T}$. The measurement noise profile $\sigma_{v,k}^2$ is plotted in  Fig. \ref{fig:3}. We set $\mu=02$ for the proposed algorithm, however, to compare the MSD of non-cooperative scheme with the proposed algorithm, we need to replace $\mu$ with $N \mu$ in \eqref{nocoop}. The MSD learning curves  for non-cooperative scheme \eqref{nocoop} and the proposed algorithm have been plotted in Fig. \ref{fig:4}. All the graphs were produced by averaging 100 independent trials. 
For the proposed algorithm, we have averaged MSD over all of the nodes, i.e. 
\begin{align}
\mathrm{MSD}=\frac{1}{N}\sum_{k=1}^N \mathrm{MSD}_k 
\end{align}
We observe that the proposed algorithm provides smaller MSD than the non-cooperative solution for all kind of the test signals.  

\begin{figure}[t]
\centering 
\includegraphics [width=8.5cm]{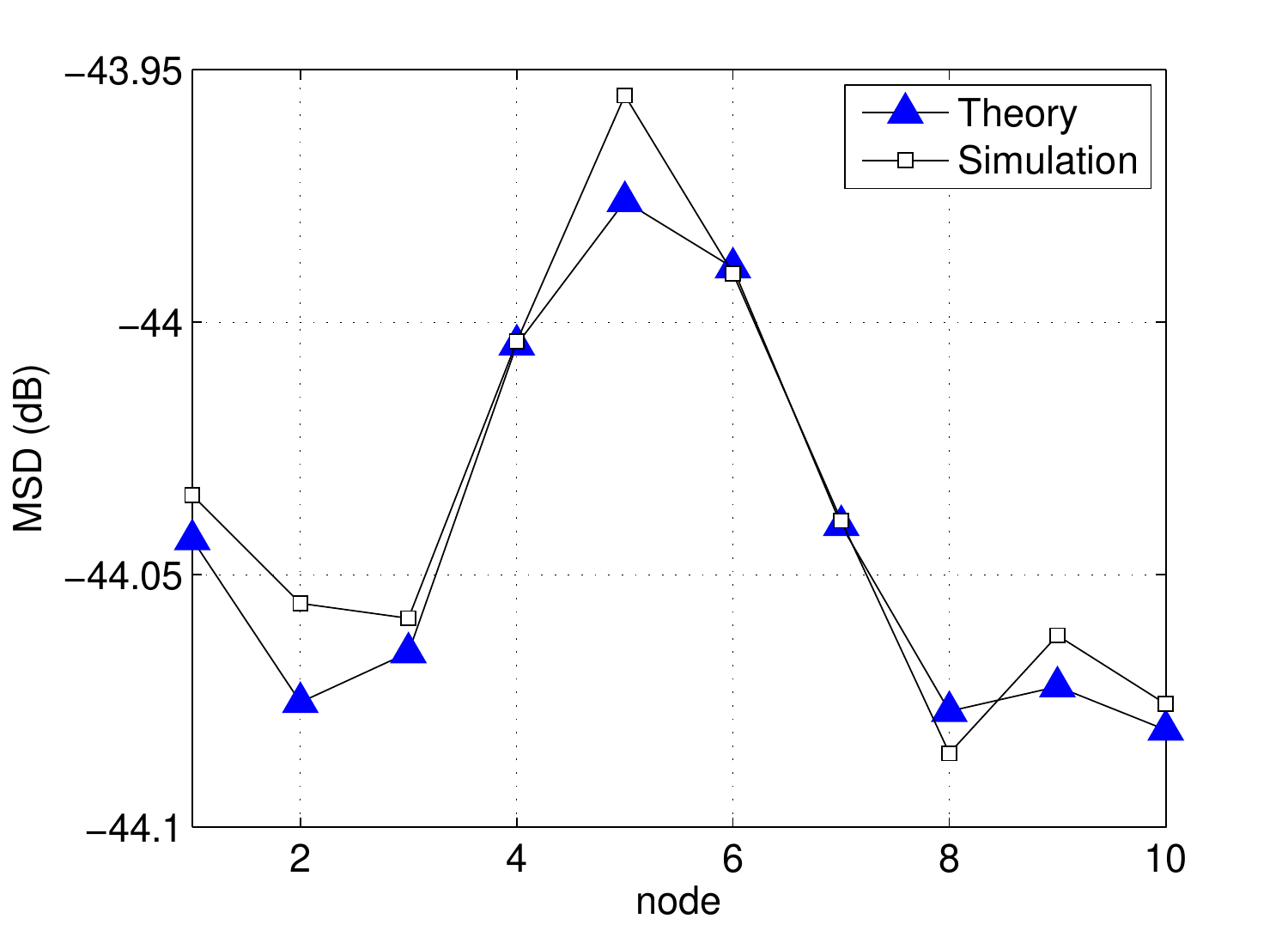} 
\centering \caption{The theoretical and simulation result values provided by the proposed at different nodes.}
\label{fig:5}
\end{figure}

\begin{figure}[t]
\centering 
\includegraphics [width=8.5cm]{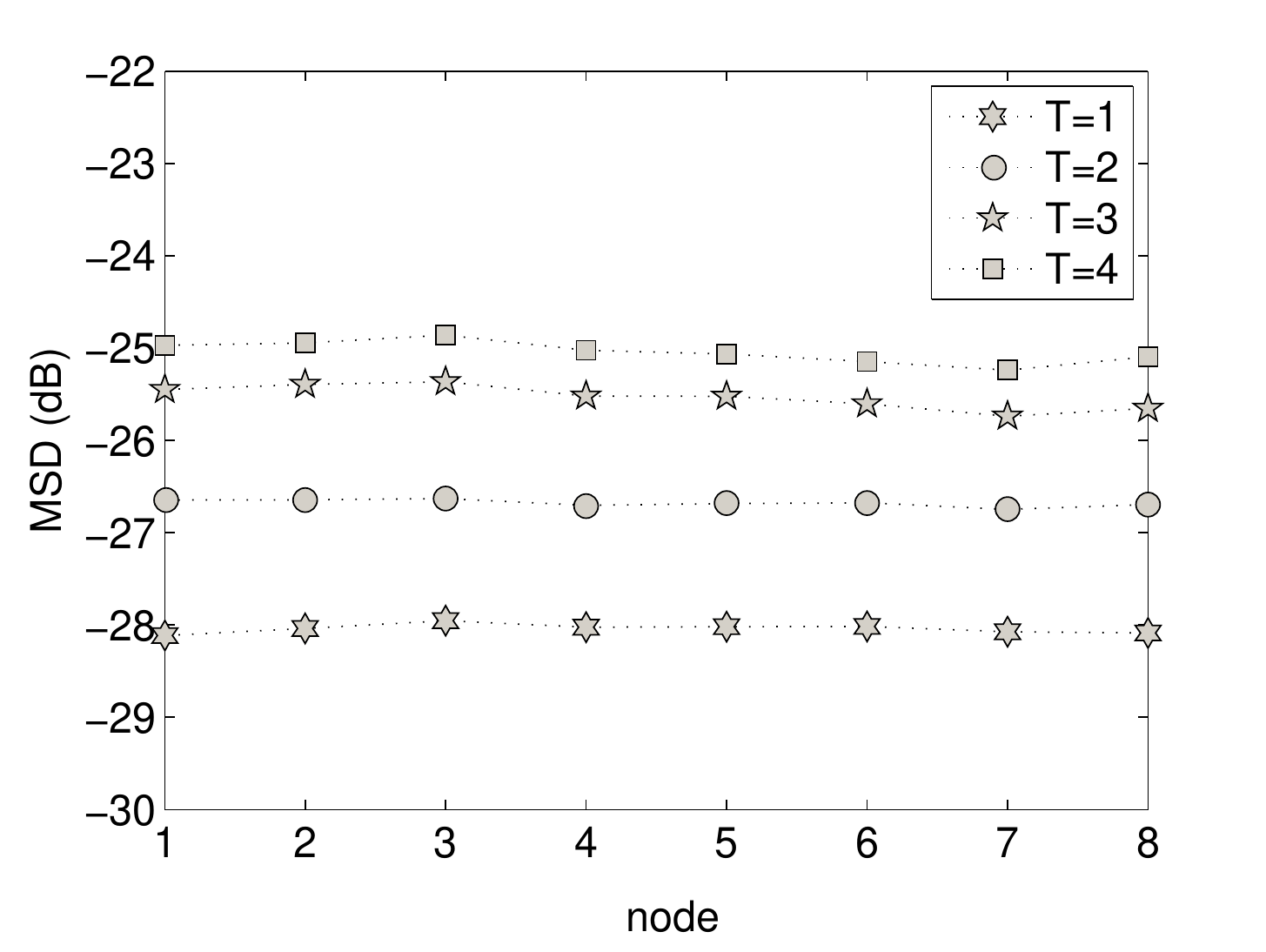} 
\centering \caption{The steady-state MSD values of proposed algorithm at every node $k$ for different values of $T$.}
\label{fig:6}
\end{figure}

In the next experiment we compare the theoretical and simulation result values, which are provided by the proposed algorithm at individual nodes for the synthesized noncircular complex-valued  signal. The steady-state values are obtained by averaging over 100 independent trials and over 50 time samples after convergence. Fig. \ref{fig:5} shows the numerical results for both simulation and the theoretical expression. We can observe that the simulation results match well the theoretical expressions. 
Fig. \ref{fig:6} shows the steady-state MSD of proposed algorithm for different values of $T$. We observe that increasing $T$ leads to a large steady-state MSD, while at the same time improves the convergence speed.

\section{Conclusion}\label{sec:5}
In this paper we proposed an adaptive estimation algorithm for in-network
processing of complex signals over distributed networks. In the proposed algorithm (incAAPA), nodes cooperate to exploit the spatio-temporal diversity; while  the affine projection learning rules enables the network (algorithm) to process  proper and improper signals. We
 extracted closed-form expressions that show how the proposed algorithm performs
in the steady-state. We further have derived the required conditions for mean-square stability. Simulation results showed validity of the theoretical results and the good performance of the proposed algorithm.

\section*{Appendix A}
If we equate the weighted energies of both sides of \eqref{wtildup}, we arrive the following space-time version of the weighted energy conservation relation for IncAAPA as:
\begin{equation} \label{eq:24}
\| \tilde{\bw}_{k,i}\|_{\bSigma_{k}}^2+\be^{\bSigma_{k}\H}_{a,k}\bF_{k,i}^{-1}\be^{\bSigma_{k}}_{a,k}=
\| \tilde{\bw}_{k-1,i}\|_{\bSigma_{k}}^2+\be^{\bSigma_{k}}_{p,k}\bF_{k,i}^{-1} \be^{\bSigma_{k}}_{p,k}
\end{equation}
Substituting \eqref{pvera} into \eqref{eq:24} and rearranging the result, we obtain
\begin{equation} \label{eq:25}
\| \tilde{\bw}_{k,i}\|_{\bSigma_{k}}^2= \| \tilde{\bw}_{k-1,i}\|_{\bSigma_{k}}^2 -\mu_k \be^{\bSigma_{k}\H}_{a,k}\bB_{k,i} \be_{k,i} 
-\mu_k \be_{k,i}^{\H} \bB_{k,i} \be^{\bSigma_{k}}_{a,k}+\mu_k^2 \be_{k,i}^{\H} \bB_{k,i} \bF_{k,i} \bB_{k,i} \be_{k,i}
\end{equation}
By using the error signal $\be_{k,i}=\bU_{k,i}^{\H} \tilde{\bw}_{k-1,i}+\bv_{k,i}$ we have
\begin{align} \label{32}
\| \tilde{\bw}_{k,i}\|_{\bSigma_{k}}^2 &=\| \tilde{\bw}_{k-1,i}\|_{\bSigma_{k}}^2-\mu_k \tilde{\bw}_{k-1,i}^{\H}\bSigma_{k}\bU_{k,i}\bB_{k,i}(\bU_{k,i}^{\H}\tilde{\bw}_{k-1,i}+\bv_{k,i})   \nonumber  \\
& -\mu_k (\tilde{\bw}_{k-1,i}^{\H}\bU_{k,i}+\bv_{k,i}^{\H})\bB_{k,i}\bU_{k,i}^{\H}\bSigma_{k}\tilde{\bw}_{k-1,i}  \nonumber  \\
& +\mu_k^2 (\tilde{\bw}_{k-1,i}^{\H}\bU_{k,i}+\bv_{k,i}^{\H})\bB_{k,i}\bF_{k,i}\bB_{k,i}(\bU_{k,i}^{\H}\tilde{\bw}_{k-1,i}+\bv_{k,i})
\end{align} 
Taking expectations of both sides of \eqref{32} and applying the Assumptions. \ref{ass:1} and \ref{ass:2} we obtain
\begin{align} \label{33}
\E \big[\|\tilde{\bw}_{k,i}\|_{\bSigma_{k}}^2\big] &=\E \big[\|\tilde{\bw}_{k-1,i}\|_{\bSigma'_{k}}^2\big]+\mu_k^2 \E\big[\bv_{k,i}^{\H}\bB_{k,i}\bF_{k,i}\bB_{k,i}\bv_{k,i}\big]
\end{align} 
where
\begin{align} \label{34}
\bSigma'_k&=\bSigma_k-\mu_k\E \big[\bSigma_k \bU_{k,i} \bB_{k,i} \bU_{k,i}^{\H}\big]
-\mu_k \E \big[\bU_{k,i} \bB_{k,i} \bU_{k,i}^{\H} \bSigma_k \big]+\mu_k^2 \E \big[\bU_{k,i} \bB_{k,i} \bU_{k,i}^{\H} \bSigma_k \bU_{k,i}\bB_{k,i}\bU_{k,i}^{\H}\big]
\end{align} 
Then, \eqref{33} and \eqref{34} can be rewritten as 
\begin{align}       \label{36}
\E \big[\|\tilde{\bw}_{k,i}\|_{\bSigma_{k}}^2\big] &=\E \big[\|\tilde{\bw}_{k-1,i}\|_{\bSigma'_{k}}^2\big]
+\mu_k^2 \E\big[\bv_{k,i}^{\H}\bD_{k,i}\bSigma_{k}\bD_{k,i}\bv_{k,i}\big]   \\
\bSigma'_k &=\bSigma_k-\mu_k \bSigma_k \E [\bD_{k,i}] \mu_k \E [\bD_{k,i} \bSigma_k]+\mu_k^2 \E [\bD_{k,i} \bSigma_k \bD_{k,i}]
\end{align} 
and the proof is complete.

\section*{References}


\end{document}